\documentclass{ws-ijmpa}
\usepackage[super,compress]{cite}
\usepackage{graphicx}

\begin{document}
\markboth{SANG PYO KIM}{Schwinger Effect, Hawking Radiation and Gauge-Gravity Relation}

%
\catchline{}{}{}{}{}
%

\title{Schwinger Effect, Hawking Radiation and Gauge-Gravity Relation}

\author{SANG PYO KIM}
\address{Department of Physics, Kunsan National University,
Kunsan 573-701, Korea\\ sangkim@kunsan.ac.kr}

\maketitle

\begin{history}
\received{Day Month Year}
\revised{Day Month Year}
\end{history}

\begin{abstract}
We present a unified picture for the Schwinger effect and the Hawking radiation and address the gauge-gravity relation and the dS-AdS duality issue
at the one-loop level. We propose a thermal interpretation for the Schwinger effect in an (A)dS space and in an Reissner-Nordstr\"{o}m black hole. The emission of charged particles from the near-extremal charged black hole is proportional to the Schwinger effect in an AdS and to another Schwinger effect in a Rindler space accelerated by the surface gravity.
\keywords{Schwinger Effect, Hawking Radiation, Thermal Interpretation, Gauge-Gravity Relation}
\end{abstract}

\ccode{PACS numbers:04.62.+v, 04.70.Dy, 12.20.-m}


\section{Introduction}\label{sec1}

A century ago Einstein discovered the theory of general relativity and almost a decade later Dirac discovered the relativistic theory of electrons. The general relativity predicts black holes and expanding universes while the relativistic electron theory predicts the Dirac sea and positrons. However, at quantum level the vacuum fluctuates, continuously creates particle-antiparticle pairs and subsequently annihilates them. The Schwinger effect, the pair production by a strong electric field,\cite{Schwinger} and the Hawking radiation, the emission of all kinds of particles by a black hole,\cite{Hawking} are two most prominent mechanisms for particle production. The spontaneous production of particles from a background field or spacetime is a nonperturbative quantum effect, which cannot be obtained by summing a finite number of Feynman diagrams.

\begin{figure}[b]
\begin{center}
\includegraphics[width=10.0cm]{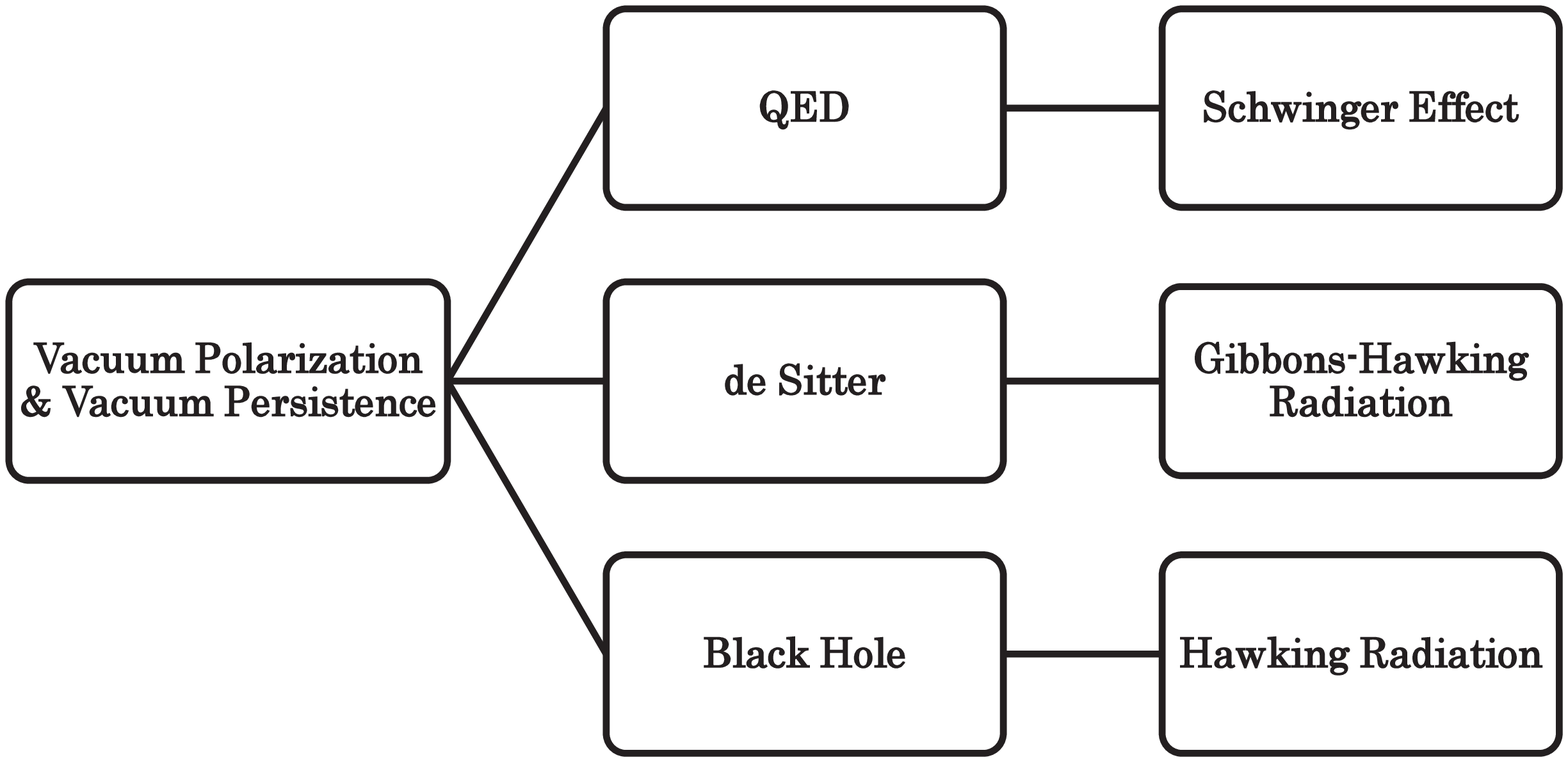}
\caption{A unified picture for the Schwinger effect, the Gibbons-Hawking radiation and the Hawking radiation due to vacuum fluctuations. \label{unified}}
\end{center}
\end{figure}
We may then raise a question of particle production mechanism when both the Schwinger effect and the Hawking radiation are simultaneously present. The first model in this direction is the Schwinger effect in a de Sitter (dS) space. The dS space emits the Gibbons-Hawking radiation with the Hawking temperature of Hubble constant.\cite{Gibbons-Hawking} Thus, a uniform electric field in the dS space exhibits both the Schwinger effect and the Hawking radiation.\cite{Garriga,Kim-Page08,Kim-Hwang-Wang,Haouat-Chekireb,FGKSSTV,Kim14a,Cai-Kim,Haouat-Chekireb15} The Schwinger effect in an anti-de Sitter (AdS) space also shows the intertwinement of quantum electrodynamics (QED) and gravity.\cite{Pioline-Troost,Kim-Page08,Cai-Kim}
The second model is the emission of particles from a charged black hole. A charged Reissner-Nordstr\"{o}m (RN) black hole emits the Hawking radiation of all species of particles due to the event horizon and also the Schwinger emission of charged particles due to the electric field on the horizon.\cite{Chen12,Kim14b,Chen14,Kim-Lee-Yoon} The charged black hole thus provides an arena where one may explore the intertwinement between QED and the quantum aspect of black hole. The Hawking radiation is determined by the Hawking temperature of the surface gravity at the horizon, whose leading term takes the Boltzmann factor
\begin{eqnarray}
{\cal N}_{\rm H} = e^{- \frac{\omega}{T_{\rm H}}}, \quad T_{\rm H} = \frac{\kappa}{2 \pi}. \label{haw bol}
\end{eqnarray}
On the other hand, the Schwinger effect in a constant electric field in a flat spacetime is given by the Davies-Unruh temperature\cite{Davies,Unruh}
\begin{eqnarray}
{\cal N}_{\rm S} = e^{- \frac{\epsilon}{T_{\rm S}}}, \quad T_{\rm S} = 2 \Bigl( \frac{qE/\epsilon}{2 \pi} \Bigr) = 2 T_{\rm U}, \label{sch tem}
\end{eqnarray}
where $\epsilon$ is the effective energy.

In this paper we present a unified picture for the Schwinger effect, the Gibbons-Hawking radiation and the Hawking radiation, and then propose a thermal interpretation for the emission of charged particles by an electric field in an ${\rm (A)dS}$ space and a charged black hole as summarized in Fig. \ref{unified}. For that purpose, we study the Schwinger effect in an ${\rm (A)dS}_2$ and the emission of charged particles from a near-extremal RN black hole. We further exploit the relation between the Maxwell theory and gravity and the duality of the dS and AdS spaces at the one-loop level. In the in-out formalism by Schwinger and DeWitt \cite{DeWitt75,DeWitt03}, the vacuum persistence and polarization is investigated for the Hawking radiation and the Schwinger effect. The thermal interpretation of the Schwinger emission in the (A)dS and the RN black hole has recently been advanced by Cai and Kim.\cite{Cai-Kim,Kim-Lee-Yoon} Using the geometry ${\rm AdS}_2 \times {\rm S}^2$ near the horizon of the extremal or near-extremal black hole, the Schwinger effect has been explicitly formulated by Chen et al.\cite{Chen12,Chen14} The emission of charged particles from the near-extremal RN black hole exhibits both the Hawking radiation and the Schwinger effect.  In fact, the Schwinger effect becomes dominant for the extremal or near-extremal RN black hole because the Hawking radiation vanishes or is exponentially suppressed. The Schwinger effect in the extremal black hole has the same form as QED in the ${\rm AdS}_2$ space,\cite{Cai-Kim,Kim-Lee-Yoon} while the Hawking radiation in the near-extremal black hole is an additional Schwinger effect in a Rindler space with the surface gravity.\cite{Kim-Lee-Yoon,Gabriel-Spindel}

\section{Schwinger Effect in (A)dS}\label{sec2}

The Schwinger effect in a two-dimensional Minkowski space is given by the mean number ${\cal N}_{\rm S} = e^{- \pi m^2/qE}$ of charged particle with mass $m$ and charge $q$. The pair-production rate is the density of states times the mean number $(qE/2 \pi) {\cal N}_{\rm S}$. The density of states originates from the wave packet in the coordinate space. The Schwinger effect has a thermal interpretation (\ref{sch tem}) in terms of the Davies-Unruh temperature  for an accelerating charge by the electric field.

A charged scalar in the vector potential $A_{\mu}$ and in a curved space $g_{\mu \nu}$ obeys the Klein-Goron equation (in the Planck units of $c = \hbar = k_B = G = 1/(4 \pi \epsilon_0) = 1$)
\begin{eqnarray}
\frac{1}{\sqrt{-g}} \hat{\pi}_{\mu} \Bigl(\sqrt{-g} g^{\mu \nu}  \hat{\pi}_{\nu} \Bigr) \phi - m^2 \phi = 0, \label{KG eq}
\end{eqnarray}
where $\hat{\pi}_{\mu} = \hat{p}_{\mu} - q A_{\mu}$. The charged scalar in a constant electric field in the ${\rm dS}_2$ or ${\rm AdS}_2$ space
\begin{eqnarray}
 ds^2 &=& - dt^2 + e^{2Ht} dx^2, \quad A_1 = - \frac{E}{H} (e^{Ht} -1), \nonumber\\
 ds^2 &=& - e^{2Kx} dt^2 + dx^2, \quad A_0 = - \frac{E}{K} (e^{Hx} -1),
\end{eqnarray}
has the instanton actions\cite{Kim-Page08}
\begin{eqnarray}
{\cal S}_{\rm dS} &=& 2 \pi \Biggl\{\sqrt{\Bigl(\frac{qE}{H^2} \Bigr)^2 + \Bigl(\frac{m}{H} \Bigr)^2 - \frac{1}{4}} - \frac{qE}{H^2} \Biggr\}, \nonumber\\
{\cal S}_{\rm AdS} &=& 2 \pi \Biggl\{\frac{qE}{K^2} - \sqrt{\Bigl(\frac{qE}{K^2} \Bigr)^2 - \Bigl(\frac{m}{K} \Bigr)^2 - \frac{1}{4}}  \Biggr\}.
\end{eqnarray}
The Boltzmann factor for the Schwinger effect
\begin{eqnarray}
{\cal N}_{\rm S} = e^{- {\cal S}_{\rm (A)dS}}, \label{boltz}
\end{eqnarray}
can be written in terms of the scalar curvature ${\cal R} = 2H^2 $ or $-2K^2$ as\cite{Kim-Page08}
\begin{eqnarray}
{\cal S} = \frac{\pi m^2}{qE} \frac{2 - \frac{\cal R}{4m^2}}{1+ \sqrt{1 + \Bigl(\frac{\cal R}{2qE} \Bigr) \Bigl( \frac{m^2}{qE} - \frac{\cal R}{8 qE} \Bigr)}}.
\end{eqnarray}

Recently, Cai and Kim have shown that the accelerating charge with the Unruh temperature in the (A)dS space may have the effective temperature and its associated temperature\cite{Cai-Kim}
\begin{eqnarray}
T_{\rm CK} = T_{\rm U} + \sqrt{T_{\rm U}^2 + \frac{{\cal R}}{8 \pi^2}}, \quad \bar{T}_{\rm CK} = \frac{1}{ 8 \pi^2} \frac{|{\cal R}|}{{T}_{\rm CK}} \label{ck tem}
\end{eqnarray}
for the Schwinger effect
\begin{eqnarray}
{\cal N}_{\rm S} = e^{- \frac{\bar{m}}{T_{\rm CK}}}, \quad \bar{m} =  \sqrt{m^2 - \frac{\cal R}{8}}.
\end{eqnarray}
It is interesting to compare the temperature (\ref{ck tem}) with another effective temperature for an accelerating observer in the (A)dS  of the Davies-Unruh temperature $T_{\rm U} = a/(2 \pi)$,\cite{Narnhofer,Deser}
\begin{eqnarray}
T_{\rm eff} = \sqrt{T_{\rm U}^2 + \frac{{\cal R}}{8 \pi^2}}.
\end{eqnarray}
The factor of 2 in the Minkowski space $({\cal R} = 0)$ is the intrinsic property of the Schwinger effect as shown in eq. (\ref{sch tem}).
The exact formula for the emission of charged scalars is given by\cite{Cai-Kim}
\begin{eqnarray}
{\cal N}^{\rm sc}_{\rm (A)dS} = \frac{e^{- \frac{\bar{m}}{T_{\rm CK}}} \pm e^{- \frac{\bar{m}}{\bar{T}_{\rm CK}}}}{1 \mp e^{- \frac{\bar{m}}{\bar{T}_{\rm CK}}}},
\end{eqnarray}
where the upper (lower) sign is for the ${\rm dS_2}$ $({\rm AdS_2})$ space.

The mean number for spin-1/2 fermion production by a constant electric field in the ${\rm dS}_2$ space can be expressed in terms of the instanton actions as\cite{Haouat-Chekireb15}
\begin{eqnarray}
{N}^{\rm sp}_{\rm dS} = \frac{e^{- ({\cal S}_{\mu} - {\cal S}_{\lambda})} - e^{-2 {\cal S}_{\mu}}}{1 - e^{- 2 {\cal S}_{\mu}}} \label{fer ds}
\end{eqnarray}
where
\begin{eqnarray}
{\cal S}_{\mu} = 2 \pi \sqrt{\Bigl(\frac{qE}{H^2} \Bigr)^2 + \Bigl(\frac{m}{H} \Bigr)^2}, \quad
{\cal S}_{\lambda} = 2 \pi \frac{qE}{H^2}. \label{ds sp act}
\end{eqnarray}
Similarly, the mean number in the ${\rm AdS}_2$ space may be given by
\begin{eqnarray}
{\cal N}^{\rm sp}_{\rm AdS} = \frac{e^{- ({\cal S}_{\kappa} - {\cal S}_{\nu})} - e^{- ({\cal S}_{\kappa}+ {\cal S}_{\nu})}}{1 - e^{({\cal S}_{\kappa}+ {\cal S}_{\nu})}}, \label{fer ads}
\end{eqnarray}
where
\begin{eqnarray}
S_{\kappa} = 2 \pi \frac{qE}{K^2}, \quad
S_{\nu} = 2 \pi  \sqrt{\Bigl(\frac{qE}{K^2} \Bigr)^2 - \Bigl(\frac{m}{K} \Bigr)^2}. \label{ads sp act}
\end{eqnarray}
Note that the Breitenlohlner-Freedman (BF) bound
\begin{eqnarray}
qE <  m K
\end{eqnarray}
prohibits fermions (scalars) from being produced in the AdS space.
Then, the Schwinger effect (\ref{fer ds}) and (\ref{fer ads}) for fermions also has the thermal interpretation
\begin{eqnarray}
{\cal N}^{\rm sp}_{\rm (A)dS} = \frac{e^{- \frac{m}{T_{\rm CK}}} - e^{- \frac{m}{\bar{T}_{\rm CK}}}}{1 - e^{- \frac{m}{\bar{T}_{\rm CK}}}},
\end{eqnarray}
where ${T}_{\rm CK}$ and $\bar{T}_{\rm CK}$ are given by eq. (\ref{ck tem}) with $T_{\rm U} = (qE/m)/(2\pi)$.

Finally, we comment on the density of states for charged particles in the (A)dS. The wave packet in the coordinate space after integrating over the momentum prescribes
\begin{eqnarray}
{\cal D} (E, {\cal R}) = \frac{1}{2 \pi} \sqrt{(qE)^2  + \frac{1}{2} m^2 {\cal R}}. \label{den st}
\end{eqnarray}
 The density of states (\ref{den st}) reduces to $qE/2\pi$ in the Minkowski space (${\cal R} = 0$) and $mH/2 \pi$ in the pure dS space ($E=0$) while the density of states and the mean number vanish in the pure AdS space. Then, the rate of pair production per unit length and per unit time is given by
 \begin{eqnarray}
\frac{d^2 {\cal N}_{(A)dS}}{dtdx} = {\cal D}_{(A)dS} (E, {\cal R}) {\cal N}_{(A)dS} (E, {\cal R}).
 \end{eqnarray}

\section{Hawking Radiation from Non-extremal Charged Black Hole}\label{sec3}

The charged RN black hole with the metric
\begin{eqnarray}
ds^2 = - \Bigl(1 - \frac{2M}{r} + \frac{Q^2}{r^2} \Bigr) dt^2 + \frac{dr^2}{1 - \frac{2M}{r} + \frac{Q^2}{r^2}} + r^2 d \Omega_2^2, \label{rn bh}
\end{eqnarray}
has the outer and inner  horizons
\begin{eqnarray}
r_{+} = M + \sqrt{M^2 - Q^2}, \quad r_{-} = M - \sqrt{M^2 - Q^2}.
\end{eqnarray}
Particles are created through the horizon of a black hole, near which the quantum field for the vacuum may work in the Rindler coordinates of the form\cite{Kim07}
\begin{eqnarray}
ds^2 = - F(\rho) d\tau^2 + \frac{d\rho^2}{G(\rho)} + r^2 (\rho) d \Omega_2^2,
\label{metric}
\end{eqnarray}
where $ d \Omega_2^2$ is the metric on the unit sphere $S^2$. The equation for the spherical harmonics $Y_{lm} (\theta, \varphi)$ of a charged scalar in a Coulomb potential $A_0 (\rho)$ takes the form
\begin{eqnarray}
\Biggl[ \Bigl( \frac{\partial}{\partial t} + i q A_0 \Bigr)^2 - \frac{\sqrt{FG}}{r^2}\frac{\partial}{\partial \rho} \Bigl(r^2 \sqrt{FG} \frac{\partial}{\partial \rho} \Bigr)+ F \Bigl(\frac{(l+\frac{1}{2})^2}{r^2 }+  m^2 \Bigr) \Biggr] \phi_{lm} (t, \rho) = 0.
\end{eqnarray}
Here, we put $(l+1/2)^2$ for the angular momentum squared.

The Hamilton-Jacobi action 
\begin{eqnarray}
S_{l} = \int \frac{d \rho}{\sqrt{FG}} \sqrt{(\omega - qA_0)^2 - F \Bigl(\frac{(l+\frac{1}{2})^2}{r^2 }+  m^2 \Bigr)}, \label{rn act}
\end{eqnarray}
from the solution
\begin{eqnarray}
\phi = e^{i S_l (\rho) - i \omega \tau} Y_{lm} (\theta, \varphi), \label{wave}
\end{eqnarray}
can give the particle production as the decay rate. In the phase-integral method, the mean number is given by the leading term\cite{Kim07,Kim-Page07,Kim13}
\begin{eqnarray}
{\cal N}_l = e^{i  {\cal S}_{\Gamma_l}},
\end{eqnarray}
where ${\cal S}_{\Gamma_l} = \oint S_l $ is the action evaluated along a contour $\Gamma_l$ in the complex plane of $\rho$.
The simple pole at $\rho = 0~(r = r_+)$ recovers the Hawking radiation of charged particles from the RN black hole
\begin{eqnarray}
{\cal N}_l = e^{- \frac{\omega - q A_0(r_+) }{T_{\rm H}}}, \label{hawking rad}
\end{eqnarray}
where $T_{\rm H}$ and $q A_0(r_+)$ are the Hawking temperature and the electrostatic potential on the horizon
\begin{eqnarray}
T_{\rm H} = \frac{r_+ - r_-}{4 \pi r_+^2}, \quad A(r_+) =  \frac{Q}{r_+}.
\end{eqnarray}

\section{Schwinger Effect in Near-extremal Charged Black Hole}\label{sec4}

The charged black hole (\ref{rn bh}), under a mapping $r - Q= \epsilon \rho, t = \tau/\epsilon$ and a reparametrization $M-Q = (\epsilon B)^2/(2Q)$, has the near-horizon geometry of ${\rm AdS}_2 \times {\rm S}^2$, \cite{Chen12}
\begin{eqnarray}
ds^2 = - \frac{\rho^2 - B^2}{Q^2} d\tau^2 + \frac{Q^2}{\rho^2 - B^2} d \rho^2 + Q^2 d \Omega_2^2. \label{AdS}
\end{eqnarray}
Then, the Coulomb potential of the black hole is $A_0 = - \rho/Q$ and the electric field on the horizon is $E_{H} =1/Q$. The Hawking radiation for the near-extremal black hole is exponentially suppressed as
\begin{eqnarray}
{\cal N}_{\rm H} = e^{- \frac{\omega - qA_0 (B)}{T_{\rm H}}}, \quad T_{\rm H} = \frac{\epsilon B}{4 \pi Q^2}.
\end{eqnarray}
On the other hand, the Schwinger effect, the emission of the same kind of charges ($qQ>0$) from the near-extremal RN black hole, is the dominant channel. The leading term for the Schwinger effect can be determined by the Hamilton-Jacobi action of the solution (\ref{wave})
\begin{eqnarray}
{\cal S}_l (\rho) = \oint \frac{d \rho}{\rho^2 - B^2} \sqrt{(q \rho - \omega Q)^2 Q^2 - \Bigl(m^2 Q^2 + (l+ \frac{1}{2})^2 \Bigr) (\rho^2 - B^2)}. \label{HJ action}
\end{eqnarray}
The particle state is well defined by the wave function (\ref{wave}) for $\rho \gg 1$ near the horizon provided that the bound
\begin{eqnarray}
Q < \sqrt{\frac{(l + \frac{1}{2})^2}{q^2 - m^2}} \label{BF bound}
\end{eqnarray}
is violated. The bound (\ref{BF bound}) is the black hole analog of the BF bound in the ${\rm AdS}$ space.\cite{Cai-Kim,Pioline-Troost,Kim08} Using the phase-integral formula again, the contour integral (\ref{HJ action}) has a pair of simple poles at $\rho = \pm B$ and another simple pole at $\rho = \infty$, whose residues contribute ${\cal S}_{a}$ and ${\cal S}_{b}$, respectively, to the Schwinger formula
\begin{eqnarray}
{\cal N}_{\rm S} = e^{- ({\cal S}_{a} - {\cal S}_{b}) }, \label{pair prod}
\end{eqnarray}
where the instanton actions are
\begin{eqnarray}
{\cal S}_{a} = 2 \pi q Q, \quad {\cal S}_{b} = 2 \pi qQ \sqrt{1 - \Bigl(\frac{m}{q} \Bigr)^2 - \Bigl(\frac{l+ 1/2}{qQ} \Bigr)^2 }. \label{instanton}
\end{eqnarray}
Note that ${\cal S}_{b}$ is a non-negative instanton action for the emission of charged scalars provided that the bound (\ref{BF bound}) is violated.

The Davies-Unruh temperature for an accelerating charge on the horizon is
\begin{eqnarray}
T_{\rm U} = \frac{q}{2 \pi \bar{m} Q},
\end{eqnarray}
where $\bar{m}$ is an effective mass,
\begin{eqnarray}
\bar{m} = m \sqrt{1+ \Bigl(\frac{l+ \frac{1}{2}}{mQ}\Bigr)^2}. \label{mass}
\end{eqnarray}
Here, the angular momentum effectively adds to the mass of charged scalars in eq. (\ref{KG eq}). Then, the emission formula (\ref{pair prod}) takes the form
\begin{eqnarray}
{\cal N}_{\rm S} = e^{- \frac{\bar{m}}{T_{\rm CK}} }, \label{Boltzmann}
\end{eqnarray}
where $T_{\rm CK}$ is the effective temperature on the horizon and $\bar{T}_{\rm CK}$ is its associated temperature
\begin{eqnarray}
T_{\rm CK} = T_{\rm U} + \sqrt{T_{\rm U}^2 - \Bigl( \frac{1}{2 \pi Q} \Bigr)^2}, \quad \bar{T}_{\rm CK} = T_{\rm U} - \sqrt{T_{\rm U}^2 - \Bigl( \frac{1}{2 \pi Q} \Bigr)^2}. \label{eff tem}
\end{eqnarray}
The reason for $T_{\rm CK}$ to be called an effective temperature is as follows. Noting that ${\cal R}_{\rm AdS} = -2/Q^2$ for the metric (\ref{AdS}), the effective mass (\ref{mass}) for an $s$-wave and the effective temperature (\ref{eff tem}) take the same form as QED in a constant electric field in the ${\rm AdS}_2$ space\cite{Cai-Kim}:
\begin{eqnarray}
\bar{m}_{\rm AdS} = m \sqrt{ 1 - \frac{{\cal R}_{\rm AdS}}{8 m^2}}, \quad T_{\rm AdS} = T_{\rm U} + \sqrt{T_{\rm U}^2  + \frac{{\cal R}_{\rm AdS}}{8 \pi^2}}
\end{eqnarray}
where $T_{\rm U} = (qE/\bar{m}_{\rm AdS})/(2 \pi)$ is the Davies-Unruh temperature.

The exact formula for the Schwinger effect from the near-extremal black hole is given by\cite{Chen12,Chen14,Kim-Lee-Yoon}
\begin{eqnarray}
{\cal N}_{\rm NBH} = \Biggl( \frac{e^{- ({\cal S}_{a} - {\cal S}_{b})} - e^{- ({\cal S}_{a} + {\cal S}_{b}) }}{1 \pm e^{- ({\cal S}_{a} + {\cal S}_{b}) }} \Biggr) \Biggl( \frac{1 \mp  e^{- ({\cal S}_{c} - {\cal S}_{a}) }}{1 + e^{- ({\cal S}_{c} - {\cal S}_{b}) }} \Biggr),
\end{eqnarray}
where the upper (lower) sign is for spinless scalars (spin-1/2 fermions) and
\begin{eqnarray}
{\cal S}_c = 2 \pi \frac{\epsilon \omega}{\epsilon B} Q^2.
\end{eqnarray}
Interestingly, the mean number of emitted particles consists of three parts, the Boltzmann amplification factor, the Schwinger effect in an ${\rm AdS}_2$, and the Schwinger effect in a two-dimensional Rindler space:
\begin{eqnarray}
{\cal N}_{\rm NBH} = e^{\frac{\bar{m}}{T_{\rm CK}}} \times \underbrace{\Biggl\{ \frac{e^{- \frac{\bar{m}}{T_{\rm CK}}} - e^{- \frac{\bar{m}}{\bar{T}_{\rm CK}}}}{1 \pm e^{- \frac{\bar{m}}{\bar{T}_{\rm CK}}}} \Biggr\}}_{\rm Schwinger~Effect~in~AdS_2} \times \underbrace{\Biggl\{ \frac{e^{- \frac{\bar{m}}{T_{\rm CK}}} \Bigl(1 \mp e^{- \frac{\omega - q A(r_{\rm H})}{T_{\rm H}}} \Bigr)}{1+ e^{- \frac{\omega - q A(r_{\rm H})}{T_{\rm H}}} e^{- \frac{\bar{m}}{T_{\rm CK}}}} \Biggr\}}_{\rm Schwinger~Effect~in~Rindler~Space}. \label{sch-haw}
\end{eqnarray}
Here, $T_{\rm CK}$ and $\bar{T}_{\rm CK}$ are defined in eq. (\ref{eff tem}). Note that $T_{\rm CK}$ is not a temperature in a genuine sense of thermodynamics since it depends on the charge to mass ratio of the emitted particles but is an Unruh temperature for accelerating charges on the horizon. The Schwinger effect in the Rindler space has been first calculated by Gabriel and Spindel.\cite{Gabriel-Spindel}.

\section{Gauge-Gravity Relation and dS-AdS Duality}

One interesting issue at the one-loop level is the gauge-gravity relation and the duality between the dS space and AdS space. Kim and Page have shown that the Boltzmann factor (\ref{boltz}) for the Schwinger effect respects the duality, ${\cal R}$ being the scalar curvature. On the other hand, it was shown in ref. \refcite{Kim-Hwang-Wang} that the exact formulae for the Schwinger effect in the global coordinates of the ${\rm dS}_2$ and ${\rm AdS}_2$ spaces satisfy the duality of the form
\begin{eqnarray}
{\cal N}^{\rm sc}_{\rm dS} ({\cal R}) {N}^{\rm sc}_{\rm (A)dS} ({\cal R}) = 1, \label{KHW rel}
\end{eqnarray}
provided that the curvature ${\cal R}$ is analytically continued beyond the BF bound. We also note the gauge-gravity relation between the Maxwell scalar for the electromagnetic theory and the scalar curvature for gravity:
\begin{eqnarray}
{\cal F}_{\rm Maxwell} = \frac{E^2}{4} \Leftrightarrow {\cal R}_{\rm scalar} = 2 H^2 ~{\rm or}~ -2K^2. \label{gaug-grav}
\end{eqnarray}
This implies that a strong electromagnetic field corresponds to a weak gravity and vice versa. The dS-AdS duality also holds for fermion production
\begin{eqnarray}
{\cal N}^{\rm sp}_{\rm dS} ({\cal R}) {N}^{\rm sp}_{\rm (A)dS} ({\cal R}) = 1.
\end{eqnarray}

The gauge-gravity relation (\ref{gaug-grav}) may hold at the level of one-loop effective action. The gauge-gravity relation has been shown for scalar QED in the ${\rm (A)dS}_2$ space.\cite{Cai-Kim} Now, for spinor QED, we may apply the reconstruction conjecture of the effective action through the Mittag-Leffler's theorem
\begin{eqnarray}
- {\cal P} \int_{0}^{\infty} \frac{ds}{s} e^{-{\cal S} s} \left[ \cot s - \frac{1}{s} + \frac{s}{3} \right] \Longleftrightarrow  - i \ln (1 - e^{-\pi {\cal S}}), \label{prop spinor}
\end{eqnarray}
to the vacuum persistence, twice the imaginary part of the integrated effective action,
\begin{eqnarray}
2 {\rm Im} W^{\rm sp}_{\rm dS} &=& - \ln \bigl(1 - e^{- ({\cal S}_{\mu} - {\cal S}_{\lambda})} \bigr) + \ln \bigl(1 - e^{- 2 {\cal S}_{\mu}} \bigr), \nonumber\\
2 {\rm Im} W^{\rm sp}_{\rm AdS} &=& - \ln \bigl(1 - e^{- ({\cal S}_{\kappa} - {\cal S}_{\nu})} \bigr) + \ln \bigl(1 - e^{- ({\cal S}_{\kappa} +{\cal S}_{\nu})} \bigr).
\end{eqnarray}
Here, ${\cal P}$ denotes the principal value prescription and the proper-time integrals are regularized by subtracting the terms which correspond to renormalization of the vacuum energy and the charge.\cite{Schwinger} Then, the vacuum polarization is a sum of proper-time integrals (\ref{prop spinor}), which exhibit the gauge-gravity relation through the actions (\ref{ds sp act}) or (\ref{ads sp act}).

\section{Conclusion}\label{sec4}

We have studied the emission of charged particles by a constant electric field in an (A)dS space and from a near-extremal Reissner-Nordstr\"{o}m black hole
and discussed the gauge-gravity relation and the dS-AdS duality in the (A)dS. The charged black hole is the arena where the Schwinger effect intertwined with the Hawking radiation and where the gauge interplays with gravity. First, in the case of an extremal black hole, one expects no Hawking radiation since the Hawking temperature vanishes. The Schwinger effect is the only channel for the emission of particles, which is nothing but the Schwinger effect in the ${\rm AdS}_2$ space with the horizon electric field since the near-horizon geometry is the ${\rm AdS}_2$ space. Second, in the case of the near-extremal black hole, the emission (\ref{sch-haw}) of charged particles is dominated by the Schwinger effect and contributed by an exponentially suppressed Hawking radiation of a tiny Hawking temperature. The exponentially small contribution is the Schwinger effect with the horizon field in an accelerating frame with the surface gravity for the Unruh temperature. The emission of neutral particles is proceeded by the Hawking radiation since the Schwinger temperature vanishes for those particles.

One intriguing question is the entropy of the extremal black hole. The Hawking temperature vanishes so that the entropy from the first law of black hole thermodynamics also vanishes. However, the entropy from the Bekenstein-Hawking area-law still is a quarter of the area of the event horizon. A question is whether the temperature (\ref{eff tem}) combined with the first law may give the entropy. In fact, assuming the emission of charges with the same mass to charge ratio and by integrating the first law, we may obtain
\begin{eqnarray}
S_{\rm BH} = \frac{m}{e} \frac{A}{4}.
\end{eqnarray}
In the unrealistic evaporation of an extremal black hole keeping the condition $m = e$, the Schwinger effect may explain the entropy of the extremal black hole. However, in the standard particle model, the ratio is extremely small ($m \ll e$) and the entropy from the Schwinger effect is a tiny fraction of the black hole entropy.

Another interesting issue not treated in this paper but worthy to pursue is the particle production from a non-extremal charged black hole. One may expect that the Hawking radiation and the Schwinger effect works together in contrast to the near-extremal black hole in this paper. Neutral particles are emitted through the Hawking radiation. In the tunneling picture, pairs are produced near the horizon of black hole. The near-horizon geometry of a near-extremal black hole is ${\rm AdS}_2 \times S^2$, while the near-horizon geometry of a non-extremal black hole is ${\rm Rindler}_2 \times S^2$. The Schwinger effect (\ref{sch-haw}) is a consequence of including one pole at the inner horizon and another pole the outer horizon for the near-extremal black hole, which are adjacent to each other. However, for quantum fields in the non-extremal black hole, pair production occurs on the horizon, which becomes the Rindler space.\cite{Kim07,Kim08} The particle production from charged black hole thus should include both the effects and will be addressed in a future publication.

\section*{Acknowledgments}
The author would like to thank Rong-Gen Cai for useful discussions, and Chiang-Mei Chen, Hyun Kyu Lee, Don N. Page and Youngsung Yoon for collaborations related to this work. He was also benefitted from discussions with Abhay Ashtekar, Remo Ruffini, Misao Sasaki and She-Sheng Xue during the International Conference on Gravitation and Cosmology (ICGC) and the fourth Galileo-Xu Guangqi Meeting at Kavli Institute for Theoretical Physics China (KITPC), Chinese Academy of Sciences on May 4-8, 2015. The participation of ICGC was supported in part by Kunsan National University and KITPC. This work was supported in part by Basic Science Research Program through the National Research Foundation of Korea (NRF) funded by the Ministry of Education (NRF-2012R1A1B3002852).


\begin{thebibliography}{0}    

\bibitem{Schwinger} J. Schwinger, {\it Phys. Rev.} {\bf 82},  664 (1951).

\bibitem{Hawking} S. W. Hawking, {\it Comm. Math. Phys.}  {\bf  43},  199 (1975).

\bibitem{Gibbons-Hawking} G. W. Gibbons and S. W. Hawking, {\it Phys. Rev. D} {\bf 15}, 2738 (1977).

\bibitem{Garriga} J. Garriga,  {\it Phys. Rev. D} {\bf 49}, 6343 (1994).

\bibitem{Kim-Page08} S. P. Kim and D. N. Page, {\it Phys. Rev. D} {\bf  78}, 103517  (2008).

\bibitem{Kim-Hwang-Wang} S. P. Kim, P. W-Y. Hwang and T-C. Wang, Schwinger mechanism in ${\rm dS_2}$ and ${\rm AdS_2}$ revisited, arXiv:1112.0885.

\bibitem{Haouat-Chekireb} S. Haouat and R. Chekireb, {\it Phys. Rev. D} {\bf 87},  088501 (2013).

\bibitem{FGKSSTV} M. B. Fr\"{o}b, J. Garriga, S. Kanno, M. Sasaki, J. Soda, T. Tanaka, and A. Vilenkin, {\it  JCAP} {\bf 04} (2014) 009.

\bibitem{Kim14a} S. P. Kim, {\it Grav. Cosmol.} {\bf 20}, 193 (2014).

\bibitem{Cai-Kim} R-G. Cai and S. P. Kim, {\it JHEP} {\bf 09} (2014) 72.

\bibitem{Haouat-Chekireb15} S. Haouat and R. Chekireb, Effect of the electric field on the creation of fermions in de-Sitter space-time, arXiv:1504.08201.

\bibitem{Pioline-Troost} B. Pioline and J. Troost, {\it JHEP} {\bf 03} (2005) 043.

\bibitem{Chen12}  C.-M. Chen, S. P. Kim, I.-C. Lin, J.-R. Sun and M.-F. Wu, {\it Phys. Rev.} {\bf D 85},  124041 (2012).

\bibitem{Kim14b} S.~P.~Kim, {\it J. Korean Phys. Soc.} {\bf 65}, 907 (2014).

\bibitem{Chen14}   C.-M. Chen, J.-R. Sun, F-Y. Tang and Ping-Yen Tsai, Spinor Particle Creation in Near Extremal Reissner-Nordstr\"{o}m Black Holes, arXiv:1412.6876.

\bibitem{Kim-Lee-Yoon}  S. P. Kim, H. K. Lee and Y. Yoon, Thermal Interpretation of Schwinger Effect in Near-Extremal Reissner-Nordstr\"{o}m Black Hole, arXiv:1503.00218.

\bibitem{Davies} P. C. W. Davies, {\it J. Phys. A} {\bf 8},  609 (1975).

\bibitem{Unruh} W. G. Unruh, {\it Phys. Rev. D} {\bf 14},  870 (1976).

\bibitem{DeWitt75} B. S. DeWitt, {\it Phys. Rept.} {\bf 19},  295 (1975).

\bibitem{DeWitt03} B. S. DeWitt, {\it The Global Approach to Quantum Field Theory}, (Oxford University Press, Oxford, 2003).

\bibitem{Gabriel-Spindel} Cl. Gabriel and Ph. Spindel, {\it Ann. Phys.} {\bf 284}, 263 (2000).

\bibitem{Narnhofer} H. Narnhofer, I. Peter and W. Thirring, {\it Int. J. Mod. Phys.} {\bf B 10}, 1507  (1996).

\bibitem{Deser} S. Deser and O. Levin, {\it Class. Quantum Grav.} {\bf 14}, L163 (1997).

\bibitem{Kim07} S. P. Kim, {\it JHEP}  {\bf 11} (2007) 048.

\bibitem{Kim13} S. P. Kim, {\it Phys. Rev. D }  {\bf 88}, 044027 (2013); {\it Phys. Lett. B} {\bf 725},  500 (2013).

\bibitem{Kim-Page07} S. P. Kim and D. N. Page, {\it Phys. Rev. D} {\bf 75},  045013 (2007).

\bibitem{Kim08} S. P. Kim, {\it J. Korean Phys. Soc.} {\bf 53}, 1095 (2008).


\end{thebibliography}
\end{document}